# Tunable photochemical deposition of silver nanostructures on layered ferroelectric $CuInP_2S_6$

Fanyi Kong,[1,a] Lei Zhang,[1,a] Tianze Cong,[2] Zhiwei Wu,[1] Kun Liu,[1] Changsen Sun,[1] Lujun Pan,[2] Dawei Li[1,b]

[1] *School of Optoelectronic Engineering and Instrumentation Science, Dalian University of Technology, Dalian 116024, China*

[2] *School of Physics, Dalian University of Technology, Dalian 116024, China*

[a] Fanyi Kong and Lei Zhang contributed equally to this work.

[b] Author to whom correspondence should be addressed: dwli@dlut.edu.cn

## ABSTRACT

2D layered ferroelectric materials such as $CuInP_2S_6$ (CIPS) are promising candidates for novel and high-performance photocatalysts, owning to their ultrathin layer thickness, strong interlayer coupling, and intrinsic spontaneous polarization, while how to control the photocatalytic activity in layered CIPS remains unexplored. In this work, we report for the first time the photocatalytic activity of ferroelectric CIPS for chemical deposition of silver nanostructures (AgNSs). The results show that the shape and spatial distribution of AgNSs on CIPS are tunable by controlling layer thickness, environmental temperature, and light wavelength. The ferroelectric polarization in CIPS plays a critical role in tunable AgNS photodeposition, as evidenced by layer thickness and temperature dependence experiments. We further reveal that AgNS photodeposition process starts from the active site creation, selective nanoparticle nucleation/aggregation, to the continuous film formation. Moreover, AgNS/CIPS heterostructures prepared by photodeposition exhibit excellent resistance switching behavior and good surface enhancement Raman Scattering activity. Our findings provide new insight into the photocatalytic activity of layered ferroelectrics and offer a new material platform for advanced functional device applications in smart memristors and enhanced chemical sensors.

## I. INTRODUCTION

Two-dimensional (2D) layered materials with noncentrosymmetric structure, such as $\alpha$-In$_2$Se$_3$, SnTe, and CuInP$_2$S$_6$ (CIPS), possess unique ferroelectric properties,[1,2] which lead to a broad spectrum of novel and advanced device applications, including ferroelectric semiconductor transistors,[3,4] optoelectronic ferroelectric memories,[5] multidirectional switchable memristors,[6] and 2D ferroelectric heterostructure-based devices.[7,8] The transition metal thiophosphate CIPS is a room-temperature ferroelectric semiconductor, with a wide bandgap of ~2.9 eV and a high Curie temperature of ~315(5) K.[9] It exhibits strong layer thickness-dependent ferroelectric behavior and retains ferroelectricity even down to atomic scale,[10,11] making it a great material platform for the potential use in ferroelectric diode,[11,12] ferroelectric field-effect transistors,[8,12,13] memristors,[14-16] ferroelectric tunnel junctions,[17] and photovoltaics.[18] Recently, theoretical studies have shown that 2D ferroelectric CIPS is also suitable for photocatalysis.[19,20] Compared with conventional bulk ferroelectrics, 2D ferroelectric materials have several advantages for photocatalytic applications. They possess a large surface area, which could supply plenty of reactive sites. More importantly, the ultrathin layer thickness, strong interlayer coupling, together with the intrinsic spontaneous polarization could lead to highly efficient separation and ultrafast transfer of photogenerated charge carriers.[19,21] Although significant progress has been made on the photocatalysis of CIPS,[19,22] experimental research on how to control its photocatalytic activity has not been reported so far.

In this work, we systemically investigate the effects of intrinsic character and external stimulation (e.g., layer thickness, temperature, light wavelength, etc) on the photocatalytic properties of CIPS for chemical deposition of silver nanostructures (AgNSs) under light irradiation [Fig. 1(a)]. We found that AgNSs are generated on both the surface and the edge for thick-layer CIPS, but only formed on the top surface for multi-/few-layer ones. This behavior can be well explained by the layer thickness induced in-plane polarization and structural phase transition. Besides, full coverage of AgNSs is observed on CIPS at temperature lower than $T_c$, while negligible AgNSs is observed above $T_c$, demonstrating that the photocatalytic activity of CIPS is tunable with the temperature via a reversible ferroelectric switch. In addition, we have achieved AgNS deposition on CIPS under separate ultraviolet (UV) or visible light irradiation, revealing its excellent photocatalytic activity over a wide wavelength. To understand the AgNS photodeposition process, we performed in situ electrical and optical imaging measurements, through which a four-step AgNS growth model is proposed. Moreover, the as-prepared AgNS/CIPS heterostructures by photodeposition exhibit an interesting conductance switching behavior (on/off ratio ~ $10^2$, power consumption ~7.4 pW) and a good surface enhancement Raman Scattering (SERS) activity. This work paves the way for the development of highly tunable and reusable photocatalysts based on ultrathin 2D ferroelectrics, and provides a new material platform with potential for neuromorphic computing and enhanced chemical sensing.

## II. RESULTS AND DISCUSSION

### A. Effect of layer thickness on the photocatalytic activity of CIPS

As illustrated in Fig. 1(a), CIPS is a layered ferroelectric material, which is composed of a sulfur framework with octahedral voids filled by the cations ($Cu^+$, $In^{3+}$) and P-P paris.[23,24] CIPS flakes with various thicknesses $h$, including thick-layer (TL, $h > 100$ nm), multi-layer (ML, $20 < h < 100$ nm) and few-layer (FL, $h < 20$ nm), were prepared by mechanical exfoliation and then transferred onto the $SiO_2$ (285 nm)/Si substrate [Fig. 1(b)]. The thickness of these samples has been identified by combing optical contrast with atomic force microscope (AFM) measurement (Fig. S1 in the supplementary material). Figure 1(c) compares the Raman spectra of TL (~185 nm), ML (~24 nm) and FL (~14 nm) CIPS flakes in Fig. 1(b), which show similar Raman active modes.[25] Multi peaks are observed in each spectrum, including 101, 124 and 137 $cm^{-1}$ for the anion, 155 $cm^{-1}$ for $\delta$(S-P-P), 234, 242, 253, 269, 281 and 300 $cm^{-1}$ for $\delta$(P-S-P), 312 $cm^{-1}$ for the cation, and 377 $cm^{-1}$ for $\nu$(P-P), consistent with the previously reported results of ferroelectric CIPS.[26,27] In this work, the photocatalytic properties of CIPS are investigated by photodeposition of AgNSs under UV light irradiation in silver nitrate ($AgNO_3$) solution [Fig.1(a)].

We first investigate the effect of layer thickness on the photocatalytic activity of CIPS [Figs. 1(d-f)]. Figure 1(d) shows the time dependent evolution of AgNSs on a TL (~240 nm) CIPS flake under 254 nm

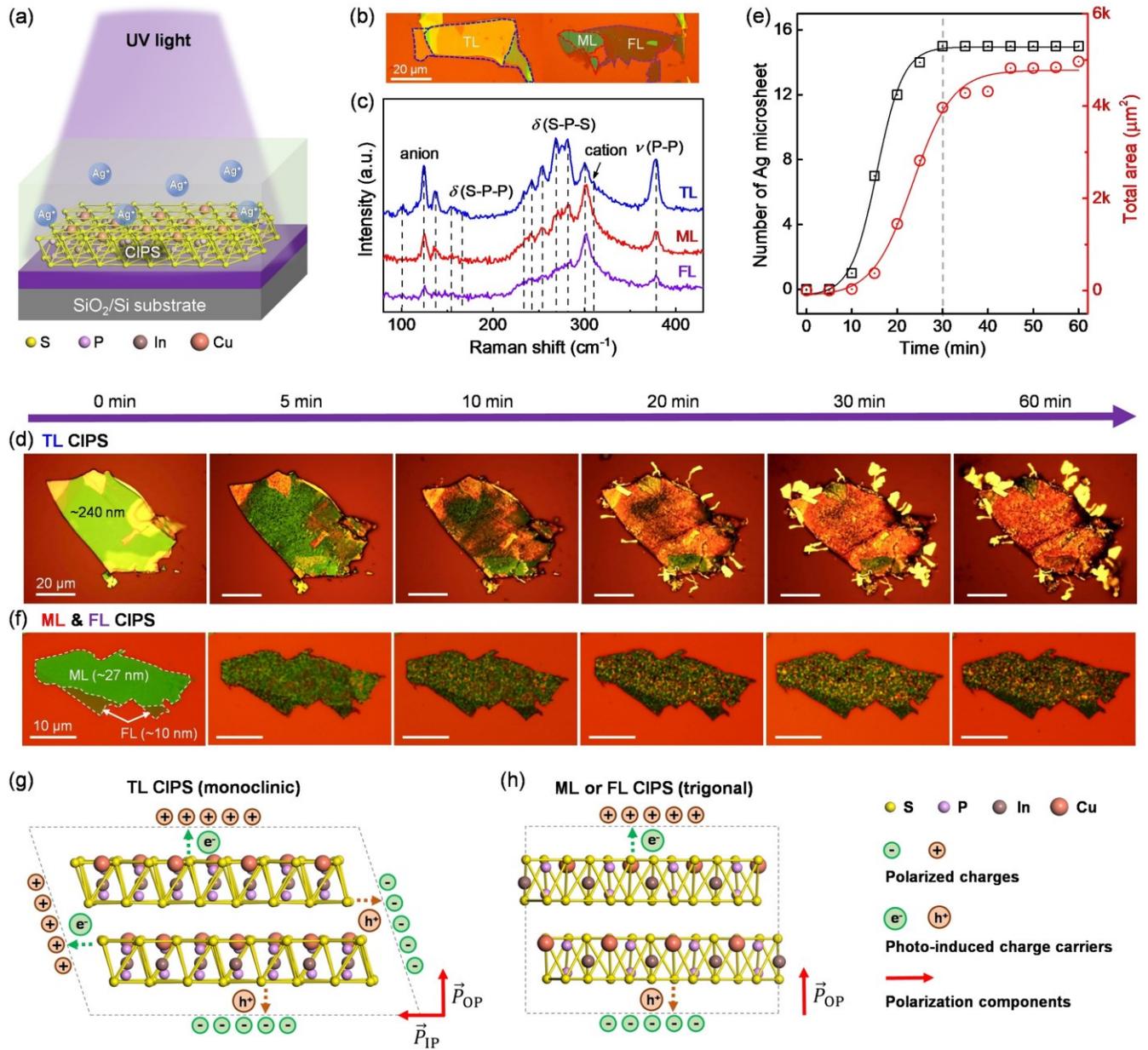

**FIG. 1.** Layer thickness dependence of photochemical deposition of silver nanostructures (AgNSs) on ferroelectric CIPS. (a) Schematic of photodeposition of AgNSs on CIPS under UV light irradiation. (b) Optical image and (c) Raman spectra for the mechanically exfoliated CIPS flakes with different thicknesses. (d, f) The effect of UV light irradiation time on AgNS formation from (d) thick-layer (TL) and (f) multi-/few-layer (ML/FL) CIPS. (e) The number (left panel) and total area (right panel) of Ag micro-sheets formed on the edge of CIPS in (d) as a function of UV irradiation time. (g, h) Schematics for (g) TL CIPS with monoclinic structure and (h) ML/FL CIPS with trigonal structure, with the corresponding polarization (red arrows) and polarization-promoted charge separation (dotted arrows).

UV light irradiation. Some key features have been found as follows [Figs. 1(d) and S2 in the supplementary material]: at the initial stage (0 < $t$ < 5 min), isolated silver nanoparticles (AgNPs) are formed on the CIPS surface, meanwhile the color of the surface changes from light green to dark green. With increasing the UV irradiation time (5 < t < 20 min), AgNPs become denser, followed by CIPS surface gradually changing into the orange color, indicating that the continuous Ag thin film has been formed in the partial region due to AgNPs aggregation. When t ≥ 20 min, Ag thin films have been deposited on most of the CIPS surface, as evidenced by the uniform distribution of orange color. In the meantime, Ag micro-sheets start to appear on the edge of CIPS and grow along in-plane direction [Fig. S2(c) in the supplementary material]. Figure 1(e) displays the Ag micro-sheet growth as a function of UV irradiation time. It can be seen that both the number and total area of Ag micro-sheets exponentially increase within 30 min and become stable with further increasing UV irradiation time. These results suggest that both the surface and the edge of TL CIPS exhibit the strong photocatalytic activity for AgNS deposition. As a comparison, we further performed AgNS photodeposition experiment on a ML (~27 nm) and a FL (~10 nm) CIPS flake under the same conditions [Figs. 1(f) and S3 in the supplementary material]. Obviously, Ag thin films have been formed on the surface, while no Ag micro-sheets is observed on the edge, revealing that the reactive sites for ML/FL CIPS only exist on its surface. Such different photocatalytic behaviors among CIPS with various thickness are confirmed by using more samples (Fig. S4 in supplementary material). Moreover, we found that after AgNSs deposited on CIPS are removed, ferroelectric CIPS can be reused without remarkable loss of its photocatalytic activity (Fig. S5 in supplementary material).

It is considered that the origin of photocatalytic activity of layered CIPS is correlated to its ferroelectric polarization characteristics. We know that the spontaneous polarization ($\vec{P}$) in ferroelectric materials could efficiently separate photogenerated electrons and holes,[28] which are then transported along the direction parallel to $\vec{P}$. In our case, CIPS exhibits a strong layer thickness-dependent structure phase transition.[29] For CIPS above a critical thickness (TL, $h > 100$ nm), it belongs to monoclinic structure and has both out-of-plane ($\vec{P}_{OP}$) and in-plane ($\vec{P}_{IP}$) polarization components [Fig. 1(g)]. Therefore, photogenerated charge carriers in TL CIPS with monoclinic structure could transport along $\vec{P}_{OP}$ and $\vec{P}_{IP}$ directions, leading to the photoreduction of AgNSs on both the surface and edge of CIPS. While for CIPS below a critical thickness (ML and FL, $h < 100$ nm), it becomes into trigonal structure and $\vec{P}_{\text{IP}}$ component disappears [Fig. 1(h)]. Under this condition, photogenerated electrons are only transported to the surface of CIPS and participate in photochemical reaction.

To understand the role of polarization (or surface charge) in the photocatalysis, we modified a CIPS flake by transferring a few-layer graphene with no polarization on top, forming graphene/CIPS heterostructure with no surface polarization (Fig. S6 in the supplementary material). We then compared the photocatalytic properties of CIPS, graphene, and graphene/CIPS heterostructure via photodeposition of AgNSs under the same condition as in Fig. 1 (Fig. S7 in the supplementary material). It can be seen that, after UV light irradiation, no obvious AgNSs is deposited in either graphene or graphene/CIPS regions, while the continuous Ag thin film together with Ag micro-sheets are formed on the surface and edge of CIPS ($t \geq 30$ min). Therefore, ferroelectric polarization plays a key role in the photocatalytic activity of layered CIPS, as will be further confirmed in the later section.

**B. In-situ monitoring of AgNS photodeposition process on CIPS**

According to the observation in Figs. 1(d) and 1(f), we propose that photodeposition of AgNSs on CIPS surface starts with the nucleation of nanoparticles, which become denser, and finally form the continuous film. To gain a better understand of this process, we performed the electrical measurement for in-situ monitoring of the evolution of AgNSs on CIPS under UV light irradiation [Fig. 2(a)]. Figure 2(b) shows the optical image of a ML (~26 nm) CIPS device, fabricated by transferring an exfoliated sample on top of two parallel Au electrodes. Current-voltage (*I-V*) curve measurement with varying voltage sweep range confirms that the pristine CIPS behaves as an insulator (Fig. S8 in the supplementary material). In-situ *I-V* testing of CIPS device was carried out during one-hour AgNS photodeposition process with a fixed sweep range from -1 to 1 V [Fig. 2(c)]. It can be seen that the conductivity of CIPS highly depends on the UV irradiation time, which changes from an insulating ($t < 10$ min) to a high conducting state ($t > 30$ min).

The current on/off ratio between high and low conductance states at $V_{bias}$ = 1 V reaches ~105 [Fig. 2(d)]. Similar phenomenon has been observed in other CIPS devices (Fig. S9 in the supplementary material).

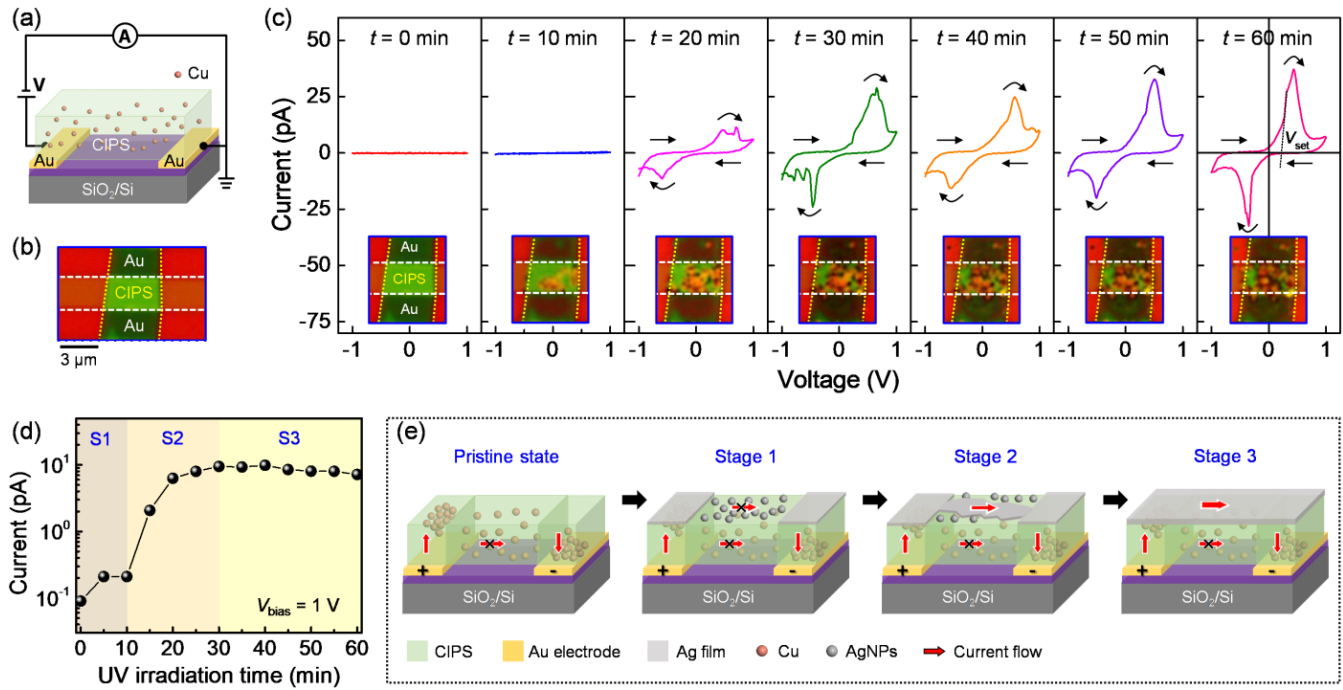

**FIG. 2.** In-situ monitoring of AgNS photodeposition on CIPS via electrical measurements. (a) CIPS device schematic. (b) Optical image of a multi-layer CIPS device, where CIPS and Au electrodes are indicated by dotted and dashed lines, respectively. (c) I-V characteristics of the CIPS device in (b) after UV light irradiation in $AgNO_3$ solution for different time, with the corresponding optical images (insets). The arrows represent voltage sweeping direction. (d) Current signal for device in (c) at a bias of 1 V as a function of UV irradiation time. (e) Modelling of three-stage AgNS photodeposition process on ferroelectric CIPS and electrical measurements.

The electrical behavior observed above definitely reflects the evolution of AgNSs on the CIPS surface [insets in Fig. 2(c)]. Based on the current versus UV irradiation time relation [Fig. 2(d)], AgNS photodeposition process on CIPS can be divided into three stages. For each stage, we proposed the possible current paths for CIPS device [Fig. 2(e)]. In the pristine state (t = 0 min), no current signal is detected. At Stage 1 (0 < t < 10 min), we observed weak current signal (~0.2 pA at $V_{bias}$ = 1 V), suggesting that isolated AgNPs have been deposited on CIPS but still not formed an effective current path. Significant current signal enhancement occurs at Stage 2 (10 < t < 30 min). During this period, the continuous Ag thin film begins to form on partial surface of the CIPS, resulting in a narrow current path. At Stage 3 ($t > 30$ min), the entire CIPS surface has been covered by Ag thin film, which makes the current channel stable.

In addition, we have observed the conductance switching (CS) behavior in AgNS/CIPS heterostructure during photodeposition, especially at Stage 3 [Figs. 2(c) and S9(b) in the supplementary material]. Such memristive effect has been previously reported in CIPS,[14-16,30] where the devices are usually designed into either lateral or vertical metal/CIPS/metal structure. The CS mechanism in CIPS is considered to be originated from the $Cu^+$ ion directional migration driven by external electric field.[14,15] In our case, no CS behavior (or current signal) was detected in the pristine CIPS, even with a large bias voltage up to 10 V (Fig. S8 in the supplementary material), thus ruling out the possibility of in-plane $Cu^+$ ions migration.[14] Instead, $Cu^+$ ions interlayer hopping movement driven by vertical electric field supports our observation and the model proposed in Fig. 2(e). After an electric field ($\vec{E}$) is applied, $Cu^+$ ions on top of the Au electrodes prefer to migrate along the $\vec{P}_{OP}$ (under positive $\vec{E}$) or -$\vec{P}_{OP}$ (under negative $\vec{E}$) direction, and then the current flowing is realized through the Ag thin film on CIPS surface. Thus, we can conclude that both directional migration of $Cu^+$ ions and Ag film on CIPS surface play the key role in the current modulation. Moreover, the power consumption ($P_{set} = I_{on} \times V_{set}$) in AgNS/CIPS heterostructure has been estimated, which is only ~7.4 pW, much lower than Au/CIPS/Ti heterostructure based device.[16] These properties make the as-prepared AgNS/CIPS heterostructures by photodeposition possible for future neuromorphic computing and smart memristors.[31-33]

Although the photodeposition of AgNSs on CIPS has been deeply analyzed via electrical measurement, this method is not effective to investigate the AgNP nucleation process [Stage 1 in Fig. 2(e)]. Thus, we performed in-situ optical imaging of the evolution of AgNSs on CIPS in the initial state [Figs. 3(a) and 3(b)]. Figure 3(b) shows a series of enlarged optical images for a CIPS flake in Fig. 3(a) exposed to UV light for different time. It can be seen that some nano textures are formed at the beginning ($t < 20$ s), which is considered to be the active sites for catalytic reaction. Indeed, nano texture positions are then gradually occupied by small AgNPs ($20 < t < 120$ s). After that, AgNPs become larger and start to emerge with the neighboring ones. Figure 3(c) shows the coverage of AgNPs extracted from Fig. 3(b) using ImageJ software as a function of UV irradiation time. It is predicted that the formation of continuous Ag thin film could be completed at $t \approx 22$ min, which is in agreement with the observation in Figs. 1 and 2. To confirm whether AgNS deposition will influence the structure of CIPS, UV irradiation time-dependent Raman measurement was performed [Fig. 3(d)]. As expected, no significant change is observed for all Raman modes of CIPS, confirming that AgNS photodeposition is a surface reaction, but not producing defects in CIPS beneath. Overall, we can conclude that the entire AgNS photodeposition process on CIPS requires four steps: (I) active site creation, (II) selective nanoparticle nucleation, (III) nanoparticle aggregation, and (IV) continuous thin film formation.

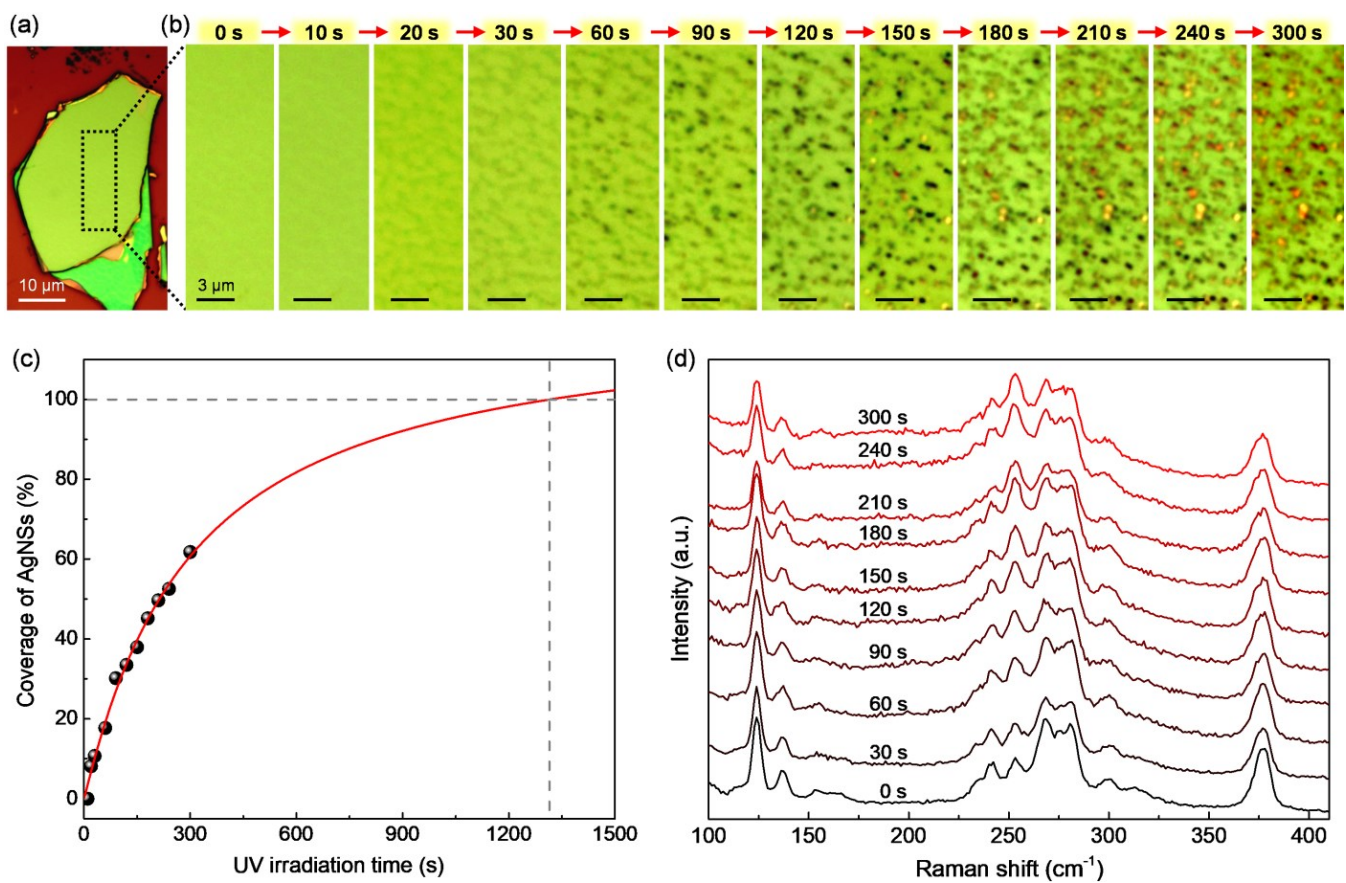

**FIG. 3.** The dynamic of the AgNS nucleation process on ferroelectric CIPS. (a) Optical image of a thick-layer CIPS flake. (b) In-situ optical imaging of the AgNS nucleation process on CIPS in (a). (c) The coverage of AgNSs on CIPS versus UV irradiation time. The dots and solid line are experimental data and fitting curve, respectively. (d) Raman spectra for the CIPS sample in (a) and (b).

## C. Effects of temperature and light wavelength on the photocatalytic activity of CIPS

To confirm the role of ferroelectric characteristic in the photocatalysis of CIPS, we performed photodeposition of AgNSs in $AgNO_3$ solution at various temperatures by choosing multiple thick-layer ($h > 100$ nm) CIPS flakes with similar ferroelectric polarization [Fig. 1(g)], where all of these samples exhibit strong room-temperature photocatalytic activity on both the surface and the edge. As shown in Fig. 4(a) and S10 in the supplementary material, at room temperature ($T = 25$ °C), Ag thin film and micro-sheets are observed on the surface and the edge of CIPS, respectively. Upon heating to $T = 45$ °C, few Ag micro-sheets are grown on the edge, and non-continuous Ag film structures are deposited on the surface. Further heating the sample to $T \geq 60$ °C, Ag micro-sheets at the edge completely disappear, while Ag film structures are grown in the center region, reflecting the influence of temperature on the distribution of reactive sites on CIPS. When the temperature is increased to 90 °C, no AgNSs is deposited on either the surface or the edge. It is believed that the transformation of CIPS from ferroelectric ($T < T_c$) to paraelectric ($T > T_c$) phase must occur during the thermal treatment mentioned above.[9,34] Figure 4(b) shows the temperature dependence of Ag film coverage on CIPS surface, demonstrating that the photocatalytic capability of CIPS for AgNS deposition in the paraelectric phase is much weaker than that in the ferroelectric phase. This is because the spontaneous polarization of CIPS, which could significantly affect the photocatalytic activities,[34,35] is lost during the ferroelectric to paraelectric phase transition, due to the movement of Cu ions within the lattice.[36] Figure 4(c) compares the room-temperature Raman spectra of CIPS before and after thermal treatment in solution at 90 °C, where no significant change is observed. It is indicated that the ferroelectric to paraelectric phase transition is a reversible switching process. In this case, the photocatalytic activity of CIPS could be gradually recovered during paraelectric to ferroelectric phase transition, as evidenced by the temperature-dependent AgNS deposition on CIPS during cooling [Figs. 4(a), 4(b), and S11 in the supplementary material]. Therefore, the photocatalytic activity of CIPS can be well tuned by ferroelectric-paraelectric phase transition, further confirming the critical role of ferroelectricity in the photocatalysis.

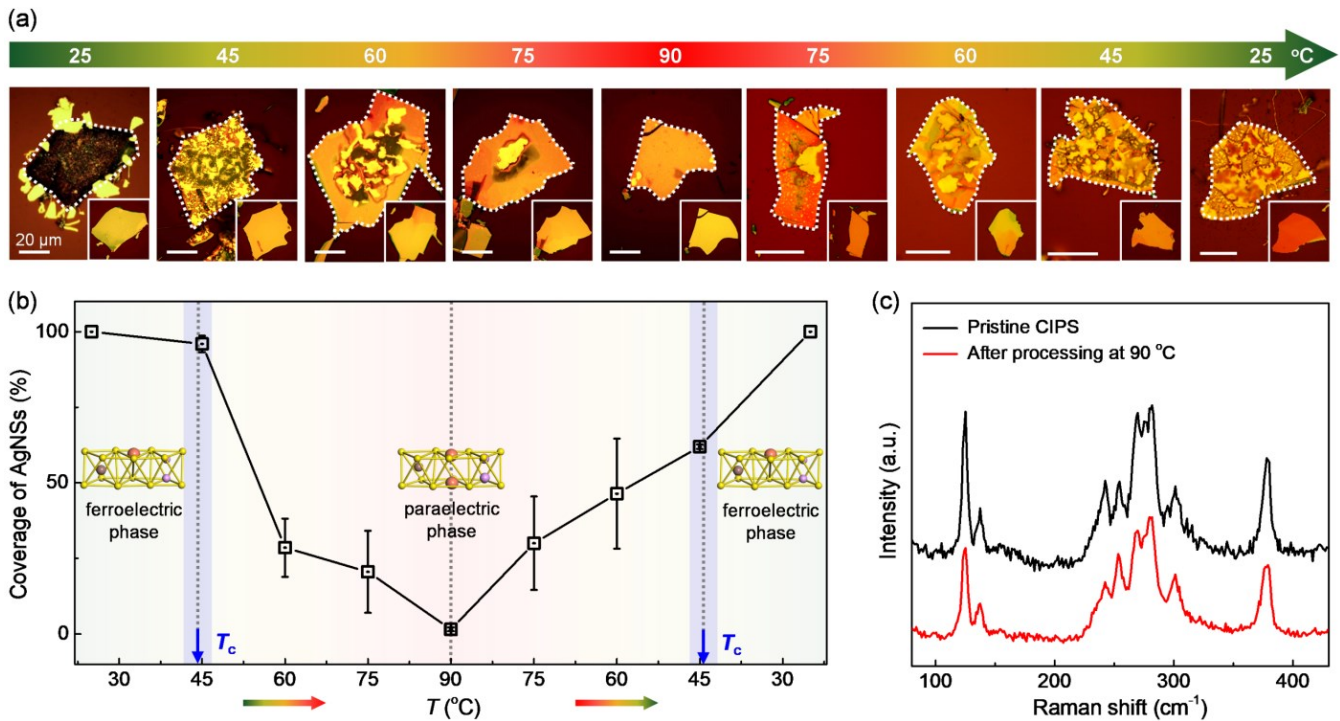

**FIG. 4.** Temperature dependence of photodeposition of AgNSs on layered CIPS. (a) Optical images of typical thick-layer CIPS flakes after UV light irradiation in AgNO₃ solution at different temperatures. Insets are the corresponding samples before reaction. The edges of CIPS flakes are marked by the white dotted lines. (b) The coverage of AgNSs on CIPS extracted from (a) versus temperature. During the thermal treatment, reversible ferroelectric-paraelectric phase transition in CIPS occurs. (c) Room-temperature Raman spectra for a CIPS sample before and after treatment in AgNO₃ solution at 90 °C.

The photocatalytic capability of CIPS is not only determined by its ferroelectricity but also highly dependent on the electronic band structure. We know that CIPS is a layered semiconductor with a wide band gap $E_g$ of ~2.9 eV. Therefore, to realize the photocatalysis of CIPS, an incident light with a photon energy above $E_g$ is required,[25] which can generate the electron-hole pair. Figures 5(a)-(c) show the optical images of TL CIPS flakes under the irradiation of light or lasers with wavelengths ranging from UV (254 nm) to visible (633 nm). Interestingly, both the UV light ($E > E_g$) and visible lasers ($E < E_g$) can be utilized to realize AgNS photodeposition on CIPS. According to semiconductor band theory [Fig. 5(d)], the origin of active photocatalysis of CIPS under UV light can be well explained by single-photon absorption (SPA). On the other hand, a unique absorption spectrum with a broad absorption tail of up to 1000 nm has been detected in CIPS,[25] which is the main reason for the observed photocatalytic activity of CIPS under the long wavelength lasers in our study. This phenomenon is considered to be a two-photon absorption (TPA) process [Fig. 5(d)],[25,37] as evidenced by photoluminescence (PL) measurement using a visible (532 nm) laser as the excitation light source [Fig. 5(e)]. We observed a strong PL spectrum with emission peak at ~2.07 eV (~600 nm), much smaller than the band gap (~2.9 eV) of CIPS, which is originated from radiative recombination at defect-states within the bandgap through interband excitation via TPA.[37]

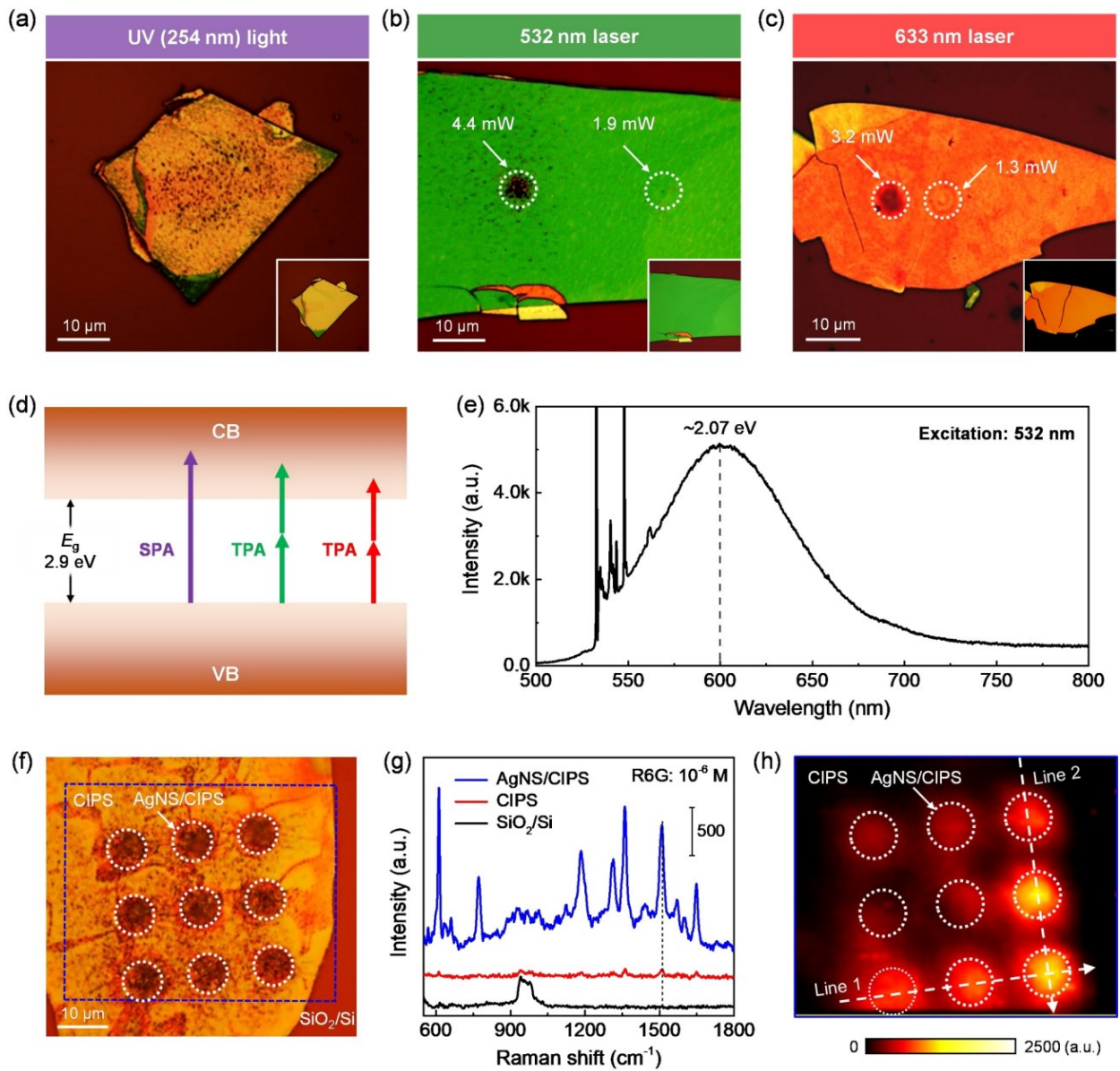

**FIG. 5.** The influence of light wavelength on the photocatalytic activity of CIPS. (a-c) Optical images of thick-layer CIPS flakes in AgNO$_3$ solution under the irradiation of (a) UV light for 5 min, (b) 532 nm laser for 4 s, and (c) 632 nm laser for 4 s. The dotted circles in (b) and (c) point to the laser irradiated areas. Insets are the corresponding samples before reaction. (d) The band structure of CIPS under light irradiation that is above (254 nm) or below (532, 633 nm) energy gap $E_g$ (~2.9 eV). (e) PL spectrum of a thick-layer CIPS under 532 nm laser excitation. (f) Optical image of the patterned AgNSs formed on CIPS by programmable laser irradiation. (g) Raman spectra of 10$^{-6}$ M R6G taken on AgNS/CIPS heterostructure, CIPS, and SiO$_2$/Si substrate. (h) 2D Raman mapping of R6G at 1508 cm$^{-1}$ measured in the boxed area in (e).

It is also noted that the location of AgNSs on CIPS could be controlled with a focused laser at an appropriate power [Figs. 5(b) and 5(c)], showing promising potential for the selective SERS detection. To prove this idea, we designed a 3×3 AgNS array by a programable 532 nm laser via photodeposition, as shown in Fig. 5(f). Figure 5(g) compares the Raman spectra of $10^{-6}$ M R6G taken at AgNS/CIPS, CIPS and $SiO_2$/Si substrate. No Raman signal of R6G is observed at $SiO_2$/Si substrate, while an enhanced Raman signal is detected at AgNS/CIPS region (Fig. S12 in the supplementary material), which is about 20 times higher than that at CIPS area. The selective SERS detection capability has been confirmed by 2D Raman mapping [Fig. 5(h)]. Although the enhancement uniformity needs further improvement, we have realized patterned SERS substrate based on ferroelectric-driven photodeposition, which is suitable for enhanced chemical sensing.

Overall, 2D ferroelectric CIPS has been proven suitable for photocatalysis, in particular, in tunable photodeposition. Compared with the conventional ferroelectrics such as perovskite $Pb(Zr,Ti)O_3$, $BaTiO_3$, and $BiFeO_3$, 2D CIPS has several advantages: (1) CIPS retains stable ferroelectricity and high photocatalytic performance even down to atomic scale (bilayers, ~1.6 nm),[11,19] which is difficult for conventional ferroelectric thin films.[38,39] (2) CIPS exhibits large intrinsic spontaneous polarization [4.22 (4.48) $\mu C/cm^2$ for OP (IP) polarization],[10] comparable with those of oxide ferroelectrics.[40] (3) A simple preparation method like exfoliation could be used to achieve highly crystalline 2D CIPS, while the fabrication of high-quality ferroelectric oxide thin films usually requires expensive equipment and careful selection of substrates with small lattice mismatch.[9] (4) CIPS is a layered structure which is flexible, but ferroelectric oxides are brittle and easy to crack, suggesting that 2D CIPS is more suitable for application in flexible ferroelectric devices such as photodetectors and photocatalysis.[41]

## III. CONCLUSION

In summary, we have achieved tunable photodeposition of AgNSs on layered CIPS by controlling the layer thickness, temperature, and light wavelength. In-situ electrical and optical imaging measurements reveal that the photodeposition of AgNSs on CIPS can be divided into four steps, including reactive site creation, selective nanoparticle nucleation, nanoparticle aggregation, and continuous film formation. Both layer thickness and temperature dependence experiments confirm that the photocatalytic activity of CIPS is much stronger in the ferroelectric phase than that in the paraelectric phase. Moreover, we found that AgNS/CIPS heterostructures prepared by photodeposition exhibit low-power conductance switching behavior and selective SERS detection capability. Thus, our work offers a new material platform for novel nanoelectronic and nanophotonic device applications, such as smart memristors, spatial selective SERS sensors, and broadband photodetectors.

## IV. EXPERIMENTAL SECTION

*Sample preparation.* Ferroelectric CIPS flakes with different layer thicknesses were mechanically exfoliated from their bulk single crystal (Six Carbon Technology, China) and then transferred onto the $SiO_2$/Si substrate. The thickness of CIPS samples was determined by combining optical microscope (ECLIPSE LV150, Nikon, Japan) and AFM (Jupiter-XR, Oxford, USA) measurements. The CIPS devices were fabricated by the standard photolithography technique. First, we prepared patterned Au (10 nm)/Ti (2 nm) electrodes on $SiO_2$/Si substrate. Second, we transferred the CIPS flake on top of the pre-patterned electrodes using our home-made 2D material dry transfer system.

*Photodeposition of AgNSs on ferroelectric CIPS.* The mechanically exfoliated CIPS samples were immersed into 3 mM $AgNO_3$ solution and exposed to UV light (wavelength: ~254 nm, power density: 18 $\mu W/cm^2$) for a fixed time. During the AgNS photodeposition process, in-situ optical and electrical monitoring experiments were carried out. In addition, we performed temperature-dependent photocatalysis experiment, where the AgNS growth was conducted under the irradiation of UV light in $AgNO_3$ solution at various temperatures ranging from 25 to 90 °C. As a comparison, the focused lasers with wavelengths of 532 and 633 nm were also chosen as the light sources. The number (or area) of AgNSs (including AgNPs and Ag thin films) deposited on top of CIPS was counted by ImageJ software.

*Characterizations.* The morphologies of CIPS and AgNS/CIPS heterostructures were characterized by optical microscope, AFM, and scanning electron microscope (SEM, JCM-500, NeoScope, Japan). The structure and composition of our samples were analyzed in a micro-Raman system (Renishaw inVia, UK). Raman signals were collected by focusing a 633 nm laser with a power of ~3.2 mW onto the sample surfaces through a ×50 objective (NA: 0.55). PL emission signals were taken with a 532 nm laser as the excitation light source. The electrical measurement was performed using a semiconductor device analyzer (B1500A, Keysight, USA) in combination with a probe station.

## SUPPLEMENTARY MATERIAL

The supplementary material contains additional information on the effect of layer thickness on the photocatalytic activity of CIPS, in-situ monitoring of AgNS deposition on CIPS, temperature-dependent photocatalytic activity of CIPS, etc.

## ACKNOELEDGMENTS

We would like to thank Xueyun Wang from Beijing Institute of Technology for CIPS model construction assistance. This work was supported by the Fundamental Research Funds for the Central Universities (No. DUT21RC(3)032, No. DUT21YG121), and the National Natural Science Foundation of China (No. 51972039).

## AUTHOR DECLRATIONS

### Conflict of Interest

The authors have no conflicts to disclose.

## DATA AVAILABILITY

The data that supports the findings of this study are available from the corresponding author upon reasonable request.